\documentclass[10pt,pra,twocolumn,showpacs,floatfix]{revtex4}

\usepackage{graphicx}
\usepackage{amssymb}
\usepackage{times}
\usepackage{amsmath}
\usepackage{amsthm}
\usepackage{epsfig}
\usepackage{epstopdf}
\usepackage{graphicx}
\usepackage{amsthm}

\makeatother
\begin{document}
\title{A Robust Interferometry Against Imperfections Base on Weak Value Amplification}
%\author
\author{Chen Fang}
\author{Jing-Zheng Huang\footnote[1]{jzhuang1983@sjtu.edu.cn}}
\author{Guihua Zeng\footnote[2]{ghzeng@sjtu.edu.cn}}

\affiliation
 {State Key Laboratory of Advanced Optical Communication Systems and Networks, Center of Quantum Information Sensing and Processing, Shanghai Jiao Tong University, Shanghai 200240, China
}

\begin{abstract}
The optical interferometry has been widely used in various high precision applications. Usually, the minimum precision of an interferometry is limited by various technique noises in practice. To suppress such kind of noises, we propose a novel scheme, which combines the weak measurement with the standard interferometry. The proposed scheme dramatically outperforms the standard interferometry in the signal noise ratio and the robustness against noises caused by the optical elements' reflections and the offset fluctuation between two paths. A proof-of-principle experiment is demonstrated to validate the amplification theory.
%The robustness of the proposed scheme indicates great potential in future applications.

%We propose the weak value amplified interferometry(WVAI) scheme for measuring small longitude phase-shift in high precision. This scheme is combining the advantages of both weak measurement and standard interferometry(SI). Compared to the SI, the WVAI scheme shows higher robustness against various practical imperfections, such as the optical elements' reflections and the offset fluctuation. Moreover, a proof-of-principle experiments is demonstrated to convince the amplification theory. Due to the similarity to SI, techniques used in SI can easily apply on WVAI with a little modification, which shows the great potentials in future applications.
\end{abstract}

\maketitle

\section{Introduction}
The optical interferometry has been widely used in science and industry fields, such as
%for hundreds years. It is based on extracting information from the phenomenon of interference caused by the superposition of waves, usually electromagnetic waves. Various applications using interferometry have been developed in
physics\cite{LIGO2016}, astronomy\cite{Astro2003}, engineering\cite{Biegen1988}, applied science\cite{atom1997}, biology\cite{DIC1981} and medicine\cite{Huang1991}. In all of these applications, an important issue is to detect the length difference or the phase difference between different paths. Making use of the obtained differences, one may achieve a highest precision length measurement\cite{LIGO2016}. Theoretically, the minimum measurable length difference is limited by the shot noise limit\cite{Giovannetti2011}, which is inversely proportional to the square root of the input intensity and the number of measurement events. While in practice, the technical noises may cause uncertainty that usually much higher than the theoretical limit. Hence to suppress the practical technical noises has become an important issue in  applications of the interferometry.

Aiming to the high precision detection, a technique called the weak value amplification(WVA)\cite{Aharonov1988,Hosten2008,Dixon2009,Viza2013,Xu2013,Fang2016,OAM2014,Hu2017} can suppress the technique noises to increase the signal noise ratio(SNR). This has been demonstrated in theories\cite{Jordan2014,Brunner2010,Li2011} and experiments\cite{Xu2013,Fang2016}. Physically, such kind of suppression can be achieved by amplifying the signal at the cost of decreasing the probability of detection. Due to the amplification, small changes beyond the detector resolution can even be detected \cite{Hosten2008,Dixon2009}.

%have attracted much attetion For this purpose, we concern with the weak value amplification (WVA) technique\cite{Aharonov1988}. This technique can amplify the signal and suppress some technique noises to increase the signal noise ratio(SNR). Owing to this, the WVA can detect small changes beyond the detector resolution, like spin Hall effect\cite{Hosten2008}, small beam deflection\cite{Dixon2009}. It could also be used to detect other changes\cite{Viza2013,Xu2013,Fang2016,OAM2014}, to enhance the gravitational wave detection method\cite{Hu2017}, to detect wave fuction\cite{Lundeen2011}, to observe the Hardy's parodox\cite{hardy2009}.
%And it has shown its potentials in precision detection its sensitivities beyond the detector's resolution
%In Ref.\cite{Brunner2010}, Brunner et al. showed that a small phase shift can be transferred to an amplified average frequency shift through WVA, which outperforms the standard interferometry technique if the misalignment errors are taken into account. Although this method has been experimentally demonstrated\cite{Xu2013,Fang2016}, the requirement of high resolution optical spectrum analysis (OSA) increases the detection complexity.
%However, the WVA technique was always compared with the standard interferometry(SI) but not used to enhanced it. It seems like the great potentials of the WVA is isolated with the SI.

In this paper, we propose a scheme named weak value amplified interferometry(WVAI) which merges the WVA and the standard interferometry(SI) together. By applying the WVA technique, this scheme amplifies the phase difference before detection in an interferometer. The amplification provides the robustness against technique noises. 
Based on an optical Mach-Zehnder interferometer, the performance of the proposed scheme is investigated. Then the influences of two kinds of technique noises, which are respectively caused by the reflections of the optical elements and the fluctuations of the phase offset, are studied. The results show that the WVAI scheme outperforms the SI in the SNR and in the the technique noise suppression. In addition, a proof-of-principle experiment is demonstrated to verify the phase amplification effect of the WVAI scheme.

This paper is organized as follows: In Sec. II we present a new interferometry scheme named WVAI. Then the influences of two typical technique noises, which are respectively caused by the reflections of the optical elements and the fluctuations of the offset between two paths, are investigated in Sec. III. In Sec. IV a proof-of-principle experiment is demonstrated and analyzed. Finally, Sec.V concludes this paper.
\section{Scheme Description}
\begin{figure}[!h]\center
	\resizebox{7cm}{!}{
		\includegraphics{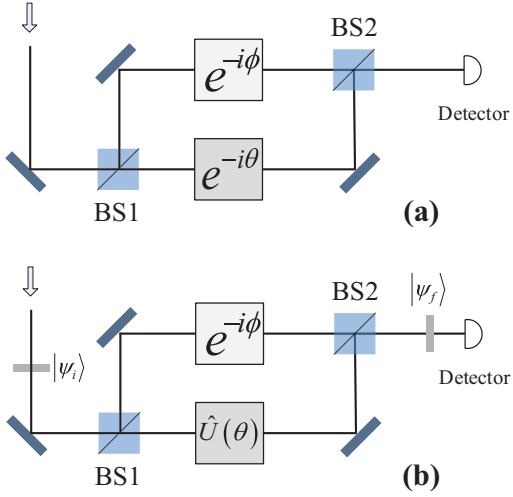}}
	\caption{(Color online). The schematic diagram of the SI(a) and the WVAI(b) based on an optical Mach-Zehnder interferometer.}\label{fig:int1}
\end{figure}

Before presenting the proposed scheme, let's consider an optical Mach-Zehnder interferometer to exemplify the standard interferometry. Displayed in Fig. \ref{fig:int1}(a), a monochromatic laser beam is split into two paths by the BS1 with 50/50 splitting ratio. In the lower path, $e^{-i\theta}$ stands for the phase difference between two paths. While in the upper path, a controllable phase $\phi$ is introduced as an offset phase delay. The two beams in the two paths interfere after they recombining by the BS2. Then one can collect the intensity of the interference by a detector, which is
\begin{equation}
\begin{aligned}
I_{D_I}(\theta) &= I_{in}|\frac{e^{-i\phi}+e^{-i\theta}}{\sqrt2}|^2\\
&=\frac{I_{in}}{4}[1+\cos(\phi-\theta)],
\end{aligned}\label{eq:IDI}
\end{equation}
where $I_{in}$ is the input light intensity. The subscript "I" represents the SI scheme.

With a little modification, one can combine the interferometer with the WVA technique. As indicated in Fig.\ref{fig:int1}(b), two linear polarizers are inserted before the BS1 and after the BS2, respectively. These polarizers are used to selected the system in the preselected state $|\psi_i\rangle=1/\sqrt{2}(|H\rangle+|V\rangle)$ and in the postselected state $|\psi_f\rangle=\cos\alpha|H\rangle+\sin\alpha|V\rangle)$, respectively. Note that H and V stands for horizontal and vertical polarized direction respectively. The phase difference $e^{-i\theta}$ is replaced with a unitary operation $\hat U(\theta)=e^{-i\theta \hat A}$, where $A\equiv|H\rangle\langle H|-|V\rangle\langle V|$. Then one obtains a WVAI scheme based on an optical Mach-Zehender interferometer. Other types of interferometer, such as Michelson, Sagnac and atom interferometer, etc., will be straightforward.

The output state of the modified interferometer is expressed as
\begin{equation}
|\Psi_{out}\rangle=\langle \psi_f|\frac{e^{-i\phi}+\hat{U}(\theta)}{\sqrt2}|\psi_i\rangle|\phi(p)\rangle,
\end{equation}
where $|\phi(p)\rangle$ is the input laser state, $p$ stands for the momentum. When $\alpha=-\pi/4+\epsilon, |\theta/\epsilon|\ll 1$, the output state could be written in its first-order-approximation
\begin{equation}
|\Psi_{out}\rangle\approx\langle \psi_f|\psi_i\rangle\frac{e^{-i\phi}+e^{-iA_w\theta}}{\sqrt2}|\phi(p)\rangle,
\end{equation}
where
\begin{equation}
A_w=\frac{\langle\psi_f|\hat A|\psi_i\rangle}{\langle\psi_f|\psi_i\rangle}=\frac{\cos\alpha-\sin\alpha}{\cos\alpha+\sin\alpha}=\cot\epsilon.
\end{equation}
The detector collects the light selected by $|\psi_f\rangle$. The detected intensity is given by
\begin{equation}
\begin{aligned}
I_{D_A}(\theta) &\approx I_{in}\langle\psi_f|\psi_i\rangle^2|\frac{e^{-i\phi}+e^{-iA_w\theta}}{\sqrt 2}|^2\\
&=\frac{I_{in}}{4A_w^2}[1+\cos(\phi-A_w\theta)],
\end{aligned}\label{eq:IDA}
\end{equation}
where the subscript "A" indicates the proposed scheme. Because of the postselection, the orthogonal polarized parts are neglected. The successful selected probability is $1/A_w^2$.

In the following, we evaluate the performance of the proposed scheme with two features. The first one is the output intensity difference between no phase delay input and $\theta(\theta>0)$ input, which is $$\Delta I_{D}=I_{D}(\theta)-I_{D}(0).$$ This difference can be obtained by differential detection\cite{Diff} or phase and amplitude modulation\cite{FOG}. The second one is the intensity contrast ratio, i.e. $I(\theta)/I(0)$, which represents the ratio of the intensity variation (detected signal) induced by the phase delay\cite{Qiu2017}.
\begin{figure}[!h]\center
\resizebox{7cm}{!}{
\includegraphics{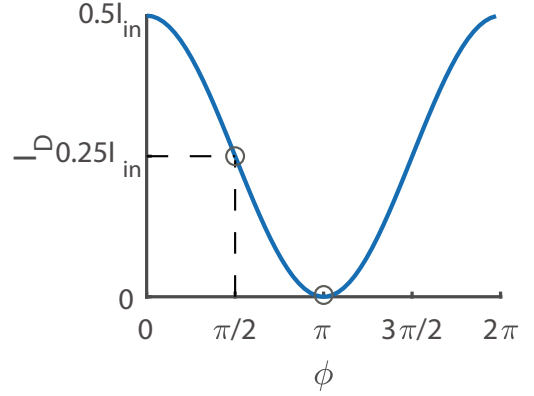}}
\caption{(Color online). The output intensity $I_D$ of different offsets $\phi$ without phase delay input in the standard interferometry.}\label{fig:offset}
\end{figure}

Fig.\ref{fig:offset} shows the output intensity of different offsets without phase delay input in the SI. Based on the WVAI scheme one may get a similar curve with an attenuation of $1/A_w^2$ in the amplitude. Typically, the offset $\phi$ is set at two phase values, i.e.  $\pi/2$ and $\pi$, for the highest sensitivity and the weak signal detection, respectively.

When $\phi=\pi$, the interferometer has total destructive interference. Since it's easier to detect a brightening of nothing than to detect a dimming of a bright light, this offset is usually set for weak signal detection like gravitational waves detection. Omitting the technique noise, $\Delta I_D$ is equal to the detected intensity in both the WVAI and the SI scheme, which can be written in the first-order approximation respectively

\begin{equation}
\begin{aligned}
  \Delta I_{D_A}=I_{D_A}\approx I_{in}\theta^2/8,\\
  \Delta I_{D_I}=I_{D_I}\approx I_{in}\theta^2/8.
  \end{aligned}
\end{equation}
Explicitly, the output intensity difference is proportional to the quadratic term of $\theta$. Although $\theta$ has been amplified, the WVAI scheme detects the same intensity as the SI does. This degradation is blamed to discarding light in the postselection. Furthermore, both methods obtain infinite intensity contrast ratios. 

\begin{figure}[!h]\center
\resizebox{7cm}{!}{
\includegraphics{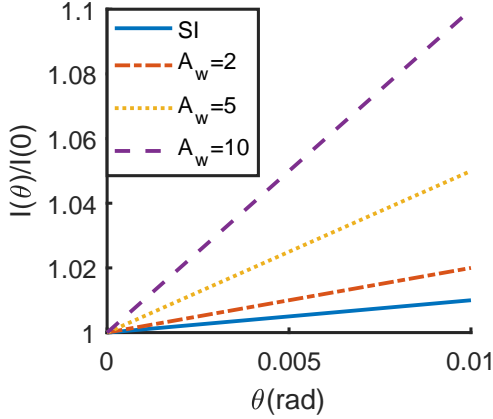}}
\caption{(Color online). The intensity contrast ratio with the offset $\phi=\pi/2$ of the SI(solid line), $A_w=2$(dash-dotted line), $A_w=5$(dotted line), $A_w=10$(dashed line).}\label{fig:INR}
\end{figure}

When $\phi=\pi/2$, this offset leads to a maximum sensitivity of the interferometer. It can be found in Fig.\ref{fig:offset}, the curve has the maximum gradient with $\phi=\pi/2$. In this situation,
\begin{equation}
\begin{aligned}
  \Delta I_{D_A}=\frac{0.25I_{in}\sin(A_w\theta)}{A_w^2}\approx\frac{0.25I_{in}\theta}{A_w},\\
  \Delta I_{D_I}=0.25I_{in}\sin\theta\approx0.25I_{in}\theta.
  \end{aligned}
\end{equation}
Obviously, $\Delta I_{D_A}$ is $1/A_w$ of $\Delta I_{D_I}$ which is caused by the low postselection's probability. However, due to the amplified phase, the WVAI outperforms than the SI on the intensity contrast ratio. As shown in Fig.\ref{fig:INR}, the intensity contrast ratio rises along with the increasing of $A_w$.
%And when $\phi$ is taking other values, WVAI's performance have a tradeoff between intensity difference and intensity contrast ratio.

Clearly, when the phase offset is set at $\phi=\pi$, the WVAI scheme and the SI scheme have the same output intensities, which means that they have the same detection uncertainty. Both scheme get infinite intensity contrast ratios. While with the offset $\pi/2$, the WVAI scheme obtain a higher intensity contrast ratio at the expense of losing the output intensity. However, if the power-recycle\cite{powerrecycle2016} is taken into consideration, all the light will go through the postselection without attenuation. One may gain a $A_w^2$ times magnificence of the output intensity, which is
\begin{equation}\
I_{D_A}(\theta)\approx \frac{I_{in}}{4}[1+\cos(\phi-A_w\theta)].
\end{equation}
The amplified phase will boost the performance with any $\phi$. With balanced differential detectors, a $\pi/2$ offset will lead to a $A_w$ times enhancement of sensitivity than SI.

\section{Imperfection analysis}
In this section, we discuss two typical kinds of technique noises as examples to demonstrate the robustness of the proposed scheme against the noises. These noises are caused by the reflections of imperfect optical elements and the fluctuation of the offset $\phi$, respectively.

\subsection{Reflection of imperfect optical elements}
%In practical experiment, the accuracy and precision of a phase-shift measurement scheme is not limited by statistics, but by the practical imperfections.

Practically, all optical elements' transmission ratio can't be 1. The reflected light will interfere with the output light field. For example, in the proposed scheme depicted in the Fig.\ref{fig:int1}, the imperfections of BS1, BS2, and all the mirrors could cause this problem. The reflected light field before postselection can be written as $E_{in}\eta_n e^{i\delta_n}$, where the subscript $n$ denotes the number of the optical elements, $E_{in}$, $\eta_n$ and $\delta_n$ are the input light field, the square root of the reflectivity, and the relative phase of corresponding elements, respectively. Generally, one has the following expression$$\eta e^{i\delta}= \sum\nolimits_{n=1}^N \eta_n e^{i\delta_n},$$where $N$ is the total number of optical elements, $\eta=\sqrt{\sum\nolimits_{n=1}^N \eta_n^2}$, and $\tan\delta=\frac{\sum\nolimits_{n=1}^N \sin\delta_n }{\sum\nolimits_{n=1}^N \cos\delta_n }$. Commonly, $\eta\ll1$ and $\delta$ can be any value from $-\pi$ to $\pi$.
Then the output intensity described by Eq. (\ref{eq:IDA}) becomes
\begin{equation}
\begin{aligned}
I_{D_A}(\theta) &\approx I_{in}\langle\psi_f|\psi_i\rangle^2|e^{-i\phi}+e^{-iAw\theta}+\sum\nolimits_{n=1}^N \eta_n e^{i\delta_n}|^2\\
&=\frac{I_{in}}{4A_w^2}\{1+\frac{\eta^2}{2}+\cos(\phi-A_w\theta)\\
&+\eta[\cos(\phi+\delta)+\cos(A_w\theta+\delta)]\}.\\
%I_{D_I}(\theta) &= I_{in}\langle\psi_f|\psi_i\rangle^2|e^{-i\phi}+e^{-i\theta}+\eta e^{i\delta}|^2\\
%&=\frac{I_{in}}{2}\{1+\frac{\eta^2}{2}+\cos(\phi-\theta)\\
%&+\eta[\cos(\phi+\delta)+\cos(\theta+\delta)]\},
\end{aligned}\label{eq:imperfect}
\end{equation}

When $\phi=\pi$, according to the Taylor expansion to the second order in Eq. (\ref{eq:imperfect}) one acquires $$\Delta I_{D_A}(\theta) \approx (S_{1A}+\sigma_{1A}^{(n)})I_{in},$$
where
\begin{equation}\left\{
\begin{aligned}
& S_{1A}=\theta^2/8\\
&\sigma_{1A}^{(n)}=\frac{\eta^2}{8A_w^2}-\eta(\frac{\theta^2}{4}\cos\delta+\frac{\theta\sin\delta}{2A_w})\\
%_{D_I}(\theta) &\approx \frac{I_{in}}{2}[\frac{\theta^2}{2}+\frac{\eta^2}{2}-\eta(\frac{\theta^2}{2}\cos\delta+\theta\sin\delta)].\\
\end{aligned}\right.
.\label{eq:noise1.A}
\end{equation}
$S_{1A}$ and $\sigma_{1A}^{(n)}$ are the proportions of the input intensity, which mean the signal induced by phase delay $\theta$ and the noise caused by the reflected light, respectively.

While in the SI, one has$$\Delta I_{D_I}(\theta) \approx (S_{1I}+\sigma_{1I}^{(n)})I_{in}$$
where
\begin{equation}\left\{
\begin{aligned}
& S_{1I}=\theta^2/8\\
&\sigma_{1I}^{(n)}=\frac{\eta^2}{8}-\eta(\frac{\theta^2}{4}\cos\delta+\frac{\theta\sin\delta}{2})\\
%_{D_I}(\theta) &\approx \frac{I_{in}}{2}[\frac{\theta^2}{2}+\frac{\eta^2}{2}-\eta(\frac{\theta^2}{2}\cos\delta+\theta\sin\delta)].\\
\end{aligned}
\right.
.\label{eq:noise1.I}
\end{equation}

\begin{figure}[!h]\center
\resizebox{9cm}{!}{
\includegraphics{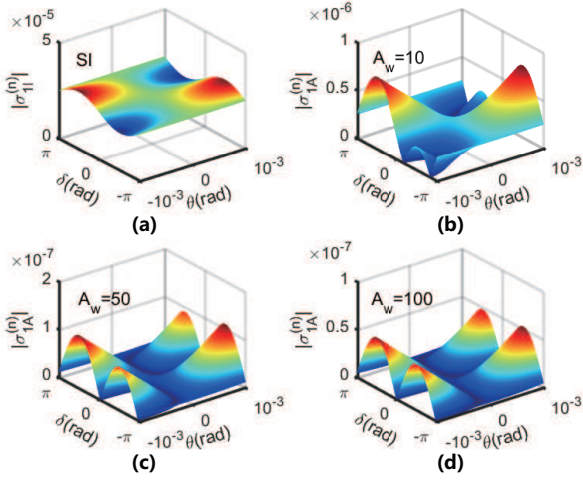}}
\caption{(Color online). The calculated results of $|\sigma_1^{(n)}|$ in Eq. (\ref{eq:noise1.A}) and Eq. (\ref{eq:noise1.I}) with $\eta=0.01$ and a offset $\phi=\pi$.}\label{fig:noise1.cos}
\end{figure}

\begin{figure}[!h]\center
\resizebox{7cm}{!}{
\includegraphics{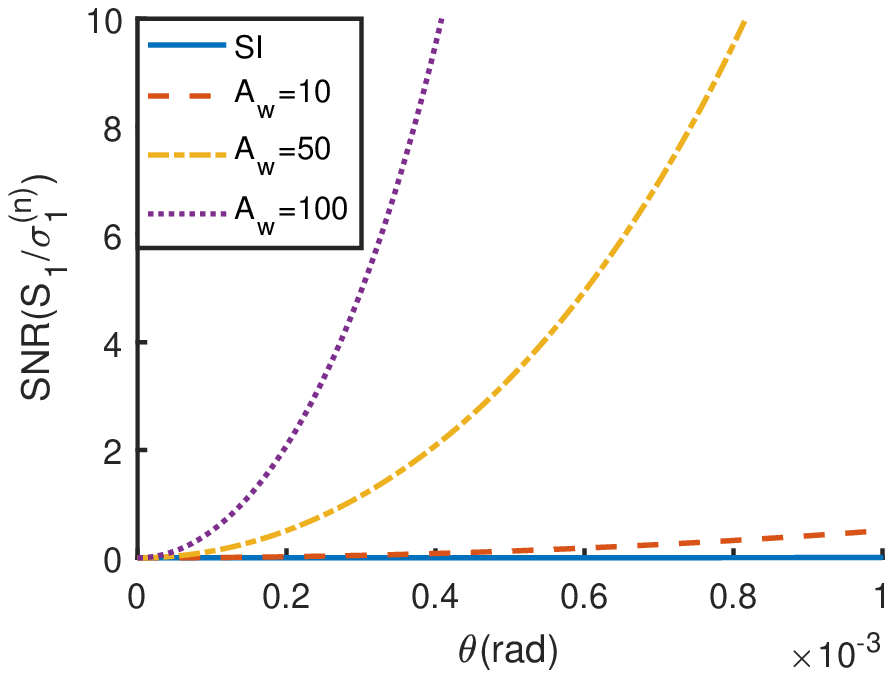}}
\caption{(Color online). The SNRs of the SI(solid line), $A_w=10$(dashed line), $A_w=50$(dash-dotted line), $A_w=100$(dotted line). They are calculated with $\eta=0.01$, $\delta=0$ and a offset $\phi=\pi$.}\label{fig:SNR1.cos}
\end{figure}

%To compare with $\eta\ll1$,
To reveal the noise influences of two ways, we compare the absolute value of $\sigma^{(n)}$ in Eq. (\ref{eq:noise1.A}) and Eq. (\ref{eq:noise1.I}) with $\eta=0.01$. The results are shown in Fig.\ref{fig:noise1.cos}. When $A_w\geq10$ the noise of the WVAI is at least 1 order of magnitude lower than the SI's, leading to a higher SNR. The SNR with $\eta=0.01$ and $\delta=0$ can been seen in Fig.\ref{fig:SNR1.cos}. In addition, compared among Fig.\ref{fig:noise1.cos}(b), (c), (d) and Fig.\ref{fig:SNR1.cos}, as long as the amplification factor raises, the noise decreases leading to increasing SNR.

When $\phi=\pi/2$, the intensity difference in the WVAI scheme is given by $$\Delta I_{D_A}(\theta) \approx (S_{2A}+\sigma_{2A}^{(n)})I_{in},$$
where
\begin{equation}\left\{
\begin{aligned}
& S_{2A}=\frac{\theta}{4A_w}\\
& \sigma_{2A}^{(n)}=\eta(\frac{1-\eta}{2A_w^2}-\frac{\theta^2}{8}\cos\delta-\frac{\theta\sin\delta}{4A_w})\\
%I_{D_I}(\theta) &\approx \frac{I_{in}}{2}[(\theta+1)-\eta(1-\eta-\frac{\theta^2}{2}\cos\delta-\theta\sin\delta)].\\
\end{aligned}\right.
.\label{eq:noise1.sin1.A}
\end{equation}

\begin{figure}[!h]\center
\resizebox{9cm}{!}{
\includegraphics{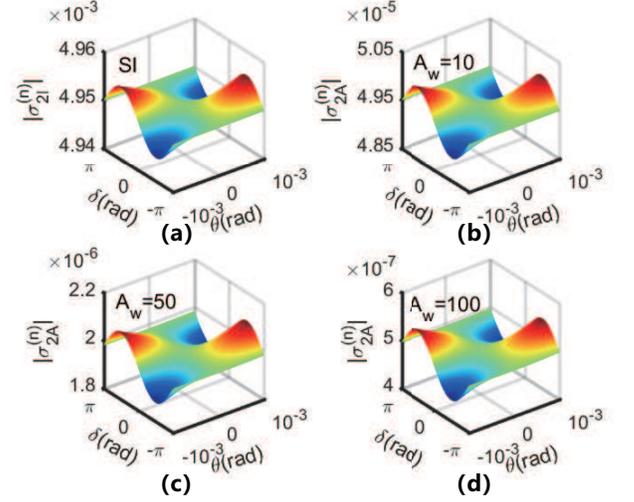}}
\caption{(Color online). The calculated results of $|\sigma_2^{(n)}|$ in Eq. (\ref{eq:noise1.sin1.A}) and Eq. (\ref{eq:noise1.sin1.I}) with $\eta=0.01$ and a offset $\phi=\pi/2$.}\label{fig:noise1.sin}
\end{figure}

\begin{figure}[!h]\center
\resizebox{7cm}{!}{
\includegraphics{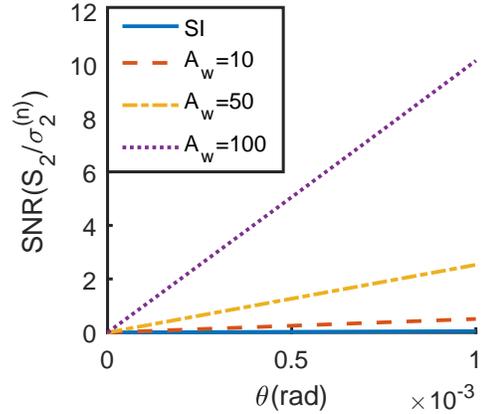}}
\caption{(Color online). The SNRs of the SI(solid line), $A_w=10$(dashed line), $A_w=50$(dash-dotted line), $A_w=100$(dotted line). They are calculated with $\eta=0.01$, $\delta=0$ and a offset $\phi=\pi/2$.}\label{fig:SNR1.sin}
\end{figure}

While in the SI scheme, the intensity difference is given by$$\Delta I_{D_I}(\theta) \approx I_{in}(S_{2I}+\sigma_{2I}^{(n)})I_{in},$$ where
\begin{equation}\left\{
\begin{aligned}
& S_{2I}=\frac{\theta}{4}\\
& \sigma_{2I}^{(n)}=\eta(\frac{1-\eta}{2}-\frac{\theta^2}{8}\cos\delta-\frac{\theta\sin\delta}{4})\\
%I_{D_I}(\theta) &\approx \frac{I_{in}}{2}[(\theta+1)-\eta(1-\eta-\frac{\theta^2}{2}\cos\delta-\theta\sin\delta)].\\
\end{aligned}\right.
.\label{eq:noise1.sin1.I}
\end{equation}
%The latter terms in the [] brackets in Eq. \ref{eq:noise1.sin1} is the noise term. $|\sigma_n|$ is
% \begin{equation}\label{eq:noise1.sin2}
%|\sigma_n|=|\eta(\frac{1-\eta}{A_w^2}-\frac{\theta^2}{2}\cos\delta-\frac{\theta\sin\delta}{A_w})|,\\
% \end{equation}
Similarly, the results of $|\sigma^{(n)}|$ with $\eta=0.01$ are depicted in Fig.\ref{fig:noise1.sin}. The noise in the WVAI scheme is at least 2 orders of magnitude less than the SI's. From Fig.\ref{fig:SNR1.sin}, although the collected intensity in the WVAI is $1/A_w$ of the SI's with the same $\theta$, the SNR of the WVAI is still higher than the SI's. %However, blamed to the attenuation of the output intensity, the SNR of the WVAI scheme reduces with a increasing $A_w$, and becomes equal to the SI's when $A_w$ is nearly to 100.
\subsection{Fluctuations of offset $\phi$}
Vibrations, air movements, deformations of optical mounts, and other environment factors may change the path difference between two arms of the interferometer. These effects cause fluctuations of $\phi$, which brings unexpected noises. Usually we use a close-loop compensation to avoid the fluctuations. However, the fluctuations can't be eliminated at all. Consider that there's a small fluctuation $\Delta\phi\ll1$. $\Delta\phi$ must be less than $\theta$, otherwise the signal will be submerged in the noise. The output intensity difference is
$$\Delta I_{D_A}(\theta)=I_{in}(S_{3A}+\sigma_{3A}^{(n)}),$$
where
\begin{equation}\left\{
\begin{aligned}
& S_{3A}=-\frac{I_{in}}{2Aw^2}\text{sin}(\phi-\frac{A_w\theta}{2})\text{sin}(\frac{A_w\theta}{2})\\
& \sigma_{3A}^{(n)}=\frac{1}{2A_w^2}\sin(\phi-A_w\theta+\frac{\Delta\phi}{2})\sin\frac{\Delta\phi}{2}
%&I_{D_I}(\theta)&=\frac{I_{in}}{2}[1+\cos(\phi+\Delta\phi-\theta)].
\end{aligned}\right.
.\label{eq:noise2.A}
\end{equation}

For the SI, one obtains$$\Delta I_{D_I}(\theta)=I_{in}(S_{3I}+\sigma_{3I}^{(n)})$$
where
\begin{equation}\left\{
\begin{aligned}
& S_{3I}=-\frac{I_{in}}{2}\text{sin}(\phi-\frac{\theta}{2})\text{sin}(\frac{\theta}{2})\\
& \sigma^{(n)}_{3I}=\frac{1}{2}\sin(\phi-\theta+\frac{\Delta\phi}{2})\sin\frac{\Delta\phi}{2}
%&I_{D_I}(\theta)&=\frac{I_{in}}{2}[1+\cos(\phi+\Delta\phi-\theta)].
\end{aligned}\right.
.\label{eq:noise2.I}
\end{equation}

\begin{figure}[!h]\center
\resizebox{7cm}{!}{
\includegraphics{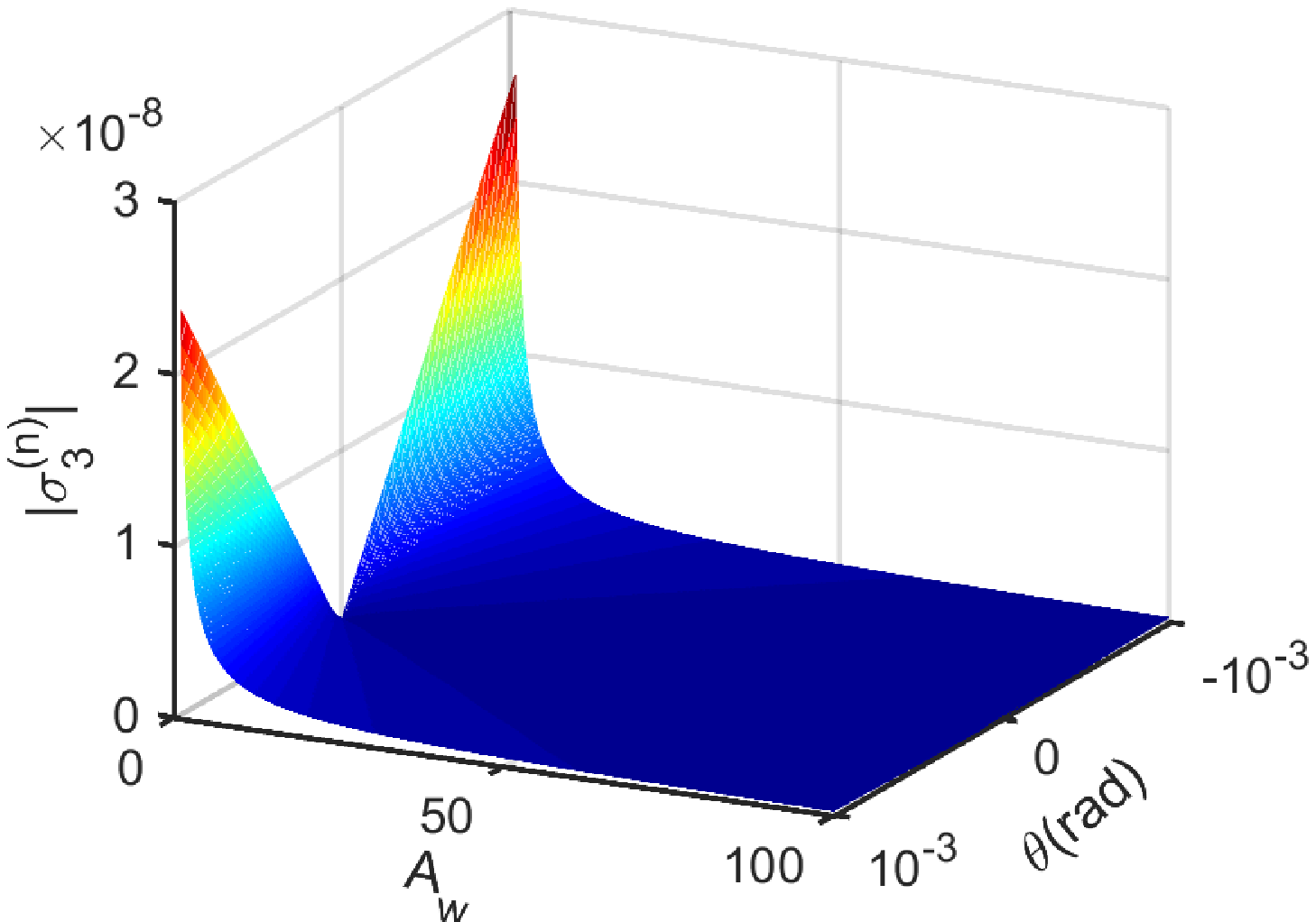}}
\caption{(Color online). The noise caused by the fluctuation of $\Delta\phi=10^{-4}$ at the offset $\pi$ calculated by Eq. (\ref{eq:noise2cos}).}\label{fig:noise2.cos}
\end{figure}

\begin{figure}[!h]\center
\resizebox{7cm}{!}{
\includegraphics{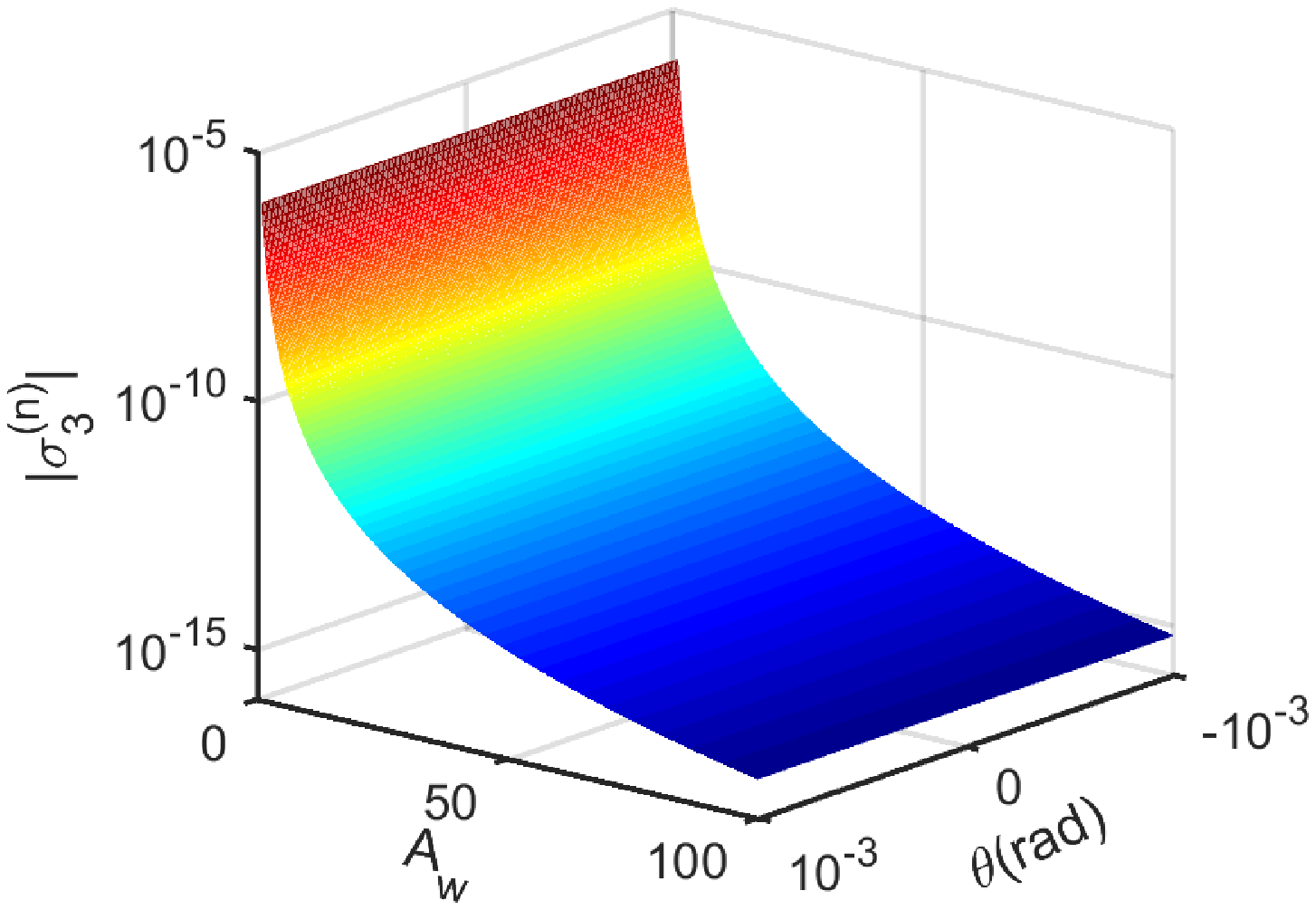}}
\caption{(Color online). The noise caused by the fluctuation of $\Delta\phi=10^{-4}$ at the offset $\pi/2$ calculated by Eq. (\ref{eq:noise2sin}).}\label{fig:noise2.sin}
\end{figure}

Obviously, when $A_w=1$,  Eq. (\ref{eq:noise2.A}) is equal to Eq. (\ref{eq:noise2.I}). Let $\phi=\pi$, $|\sigma_3^{(n)}|$ becomes
\begin{equation}\label{eq:noise2cos}
\begin{aligned}
|\sigma_3^{(n)}|&=|\frac{1}{2A_w^2}\sin(A_w\theta-\frac{\Delta\phi}{2})\sin\frac{\Delta\phi}{2}|\\
&\approx|\frac{(2A_w\theta-\Delta\phi)\Delta\phi}{8A_w^2}|.
%|\sigma_n|&\approx \frac{\Delta\phi I_{in}}{2}[\frac{1-\frac{\Delta^2\phi}{4}}{A_w^2}+\frac{\Delta\phi}{2A_w}+1]\\
\end{aligned}
\end{equation}
When $|\theta|\leq|\Delta\phi|$, $|\sigma_3^{(n)}|$ decreases monotonically along $A_w\geq1$, so the noise of WVAI scheme is less than the SI's. The SNR of the WVAI scheme could benefit from the attenuated noise. The simulated results with $\Delta\phi=10^{-4}rad$ is illustrated in Fig.\ref{fig:noise2.cos}.
Let $\phi=\pi/2$, Eq. (\ref{eq:noise2cos}) becomes
\begin{equation}\label{eq:noise2sin}
\begin{aligned}
|\sigma_3^{(n)}|&=|\frac{1}{2A_w^2}\cos(A_w\theta-\frac{\Delta\phi}{2})\sin\frac{\Delta\phi}{2}|\\
&\approx|\frac{(1-A_w^2\theta^2-\frac{\Delta^2\phi}{4}+A_w\theta)\Delta\phi}{8A_w^2}|.
%|\sigma_n|&\approx \frac{\Delta\phi I_{in}}{2}[\frac{1-\frac{\Delta^2\phi}{4}}{A_w^2}+\frac{\Delta\phi}{2A_w}+1]\\
\end{aligned}
\end{equation}
$|\sigma_3^{(n)}|$ also monotonically decreases along $A_w\geq1$ when $\phi=\pi/2$. Fig.\ref{fig:noise2.sin} is the calculated diagram with $\Delta\phi=10^{-4}rad$. Apparently, the SNR of the WVAI outperforms than the SI's with a $\pi/2$ offset.

\section{Experiment}
\begin{figure}[!h]\center
\resizebox{7cm}{!}{
\includegraphics{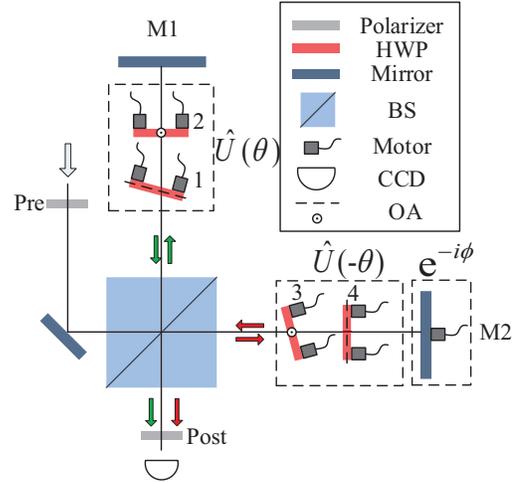}}
\caption{(Color online). The experimental setup. HWP: half wave plate. BS: 50/50beam splitter. OA: optical axis.}\label{fig:exp}
\end{figure}
As mentioned in Sec. II, the results of other types of interferometer is similar to the Mach-Zehnder type's. Because it's easier to build and adjust, we choose a Michelson interferometer to demonstrate the proposed scheme instead.
The experiment is set as illustrated in Fig.\ref{fig:exp}. A monochromatic laser beam, generated by a diode laser(Toptica, DL100) with a central wavelength of $\lambda_0=780\text{nm}$, is prepared in the state$|\psi_i\rangle=1/\sqrt2(|H\rangle+|V\rangle)$ by the first polarizer with label "Pre". Then the beam enters a Michelson interferometer. The phase difference between two arms is $\phi+2n\pi$, where $n$ is integer. The difference is set by the motor fixed on the mirror M2. After the beams recombining through BS, a polarizer with label "Post" selected them at state $|\psi_f\rangle=\cos\alpha|H\rangle+\sin\alpha|V\rangle$. In both arms of the interferometer, we place a two-waveplates-group to introduce a phase delay between horizontal and vertical polarization. It has been proved in \cite{Xu2013,Fang2016} the two-waveplates-group could equivalently realize a thin birefringent crystal. The HWPs' OAs(optical axes) are perpendicular to each other to cancel their phase delay. Tilting one of the HWPs around its OA by a tiny angle $\gamma$ increases the optical path of this HWP, which introduce a unitary operation $\hat U(\theta)=e^{-i\theta A}$. In this experiment, the HWPs are binary com-
pound zero-order half-wave plates, so the relationship between $\gamma$ and $\theta$ is\cite{Fang2016}
\begin{equation}
\begin{aligned}
\theta=\frac{\pi(n_e-n_o)h\gamma^2}{\lambda n^2},
\end{aligned}
\end{equation}
where $n_e$, $n_o$, and $n$ are refractive indices of quartz for extraordinary light, ordinary light, and average light, respectively, $h$ is the thickness of the plate, and $\lambda$ is the wavelength of the light.
The tilt also increase the optical path in one arms which cause a variation of offset $\phi$, so we put the HWPs into both arms to diminish this influence. As shown in Fig.\ref{fig:exp}, HWP1's and HWP4's OAs are horizontal direction. HWP2's and HWP3's OAs are vertical direction. We tilt the HWP1 and HWP3 along their OAs by the same angle $\gamma$ to introduce opposite phase delay and to compensate the increased optical path length.

By rotating the postselection polarizer to a different position, we could set required $A_w$. The offset is set at $\pi$ for the total destructive interference. However, the minimum step size of the motor on the mirror M2 limits the practical accuracy of setting $\phi$. The vibration of the optical platform and the deformation of the motor cause fluctuations of the offset. To prove the amplification effect, we use intensity contrast ratio $I(\theta)/I(0)$ as a criterion. When $\phi>A_w\theta$, we obtain that
\begin{equation}
\begin{aligned}
\frac{I(\theta)}{I(0)}=\frac{1+\cos(\phi-2A_w\theta)}{1+\cos\phi}.
\end{aligned}\\ \label{eq:curve1}
\end{equation}
When $|\psi_f\rangle=|H\rangle$, the experiment becomes a traditional Michelson interferometer with the path difference of $\phi-2\theta$ between two arms. In this situation the intensity contrast ratio becomes
\begin{equation}
\begin{aligned}
\frac{I(\theta)}{I(0)}=\frac{1+\cos(\phi-2\theta)}{1+\cos\phi}.
\end{aligned}\\ \label{eq:curve2}
\end{equation}
which are same as the standard interferometry's results discussed before.
\begin{figure}[!h]\center
\resizebox{9cm}{!}{
\includegraphics{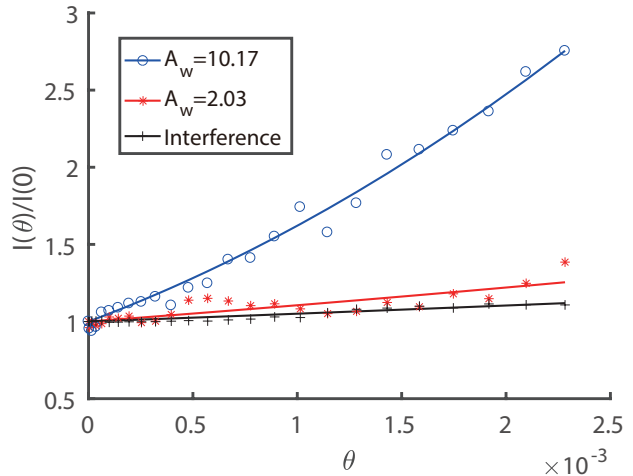}}
\caption{(Color online). Relation of the phase delay $\theta$ and intensity contrast ratio $I(\theta)/I(0)$. Blue circles, red plus signs and black asterisks are experimental results with $A_w=10.17,2.03$ and interference, respectively. And the corresponding lines are fitting with Eq. (\ref{eq:curve1}) and Eq. (\ref{eq:curve2}) .}\label{fig:result}
\end{figure}

The experimental results are shown in Fig.\ref{fig:result}. All curves fit well with Eq. (\ref{eq:curve1}) and Eq. (\ref{eq:curve2}). We calculate that $\phi=3.1639$, which meet the set value $\pi$ mentioned before. Apparently, the intensity contrast ratio increases along the amplification factor $A_w$ as predicted in Sec.II. The results firmly support the amplification effect in the WVAI scheme.
%Because of the non-isolated optical platform, the fluctuation of the laser source, and the electrical noise of CCD detector, the experiment suffers high noises. We will improve the experiment result with close-loop compensation and other noise cancelling methods as a future work.

\section{Conclusion}

In summary, we propose a new scheme named weak value amplified interferometry. The proposed scheme combines the weak value amplification and the SI. It can amplify the phase difference between different paths at a cost of decreasing the detected probability. Benefited from the amplification, the proposed scheme is robust against the technique noises. To exemplify the scheme, the investigation is based on an optical Mach-Zehnder interferometer. Comparing to the SI, the proposed scheme shows a higher intensity contrast ratio and stronger suppressions against two practical noises, which are respectively caused by the reflections of the optical elements and the fluctuations of the offset, with the offset setting at $\pi$ and $\pi/2$.
%It's worth noting that the similarity between the WVAI and the SI indicates the WVAI's potentials in analogizing the developed techniques of the SI to cover some disadvantages. For instance, using power recycle can boost the output intensity of the WVAI, and a close-loop compensation can enlarge the dynamic range. 
To demonstrate the proposed scheme, a proof-of-principle experiment based on a Michelson interferometer is established and analyzed. The results successfully validate the phase amplification effect.

\end{document}